\documentstyle[12pt,aasms4]{article}
\lefthead{Gao et al.}
\righthead{Superhumps in ER UMa}
\begin{document}

\title{Superhumps in a Peculiar SU UMa-Type Dwarf Nova\\ 
     ER Ursae Majoris}
\author{Weihong Gao, Zongyun Li, Xiaoan Wu}
\affil{Department of Astronomy, Nanjing University, Nanjing 210093, China}
\authoremail{zyli@nju.edu.cn}
\and
\author{Zhousheng Zhang, Yulan Li}
\affil{Yunnan Observatory, Kunming 650011, China}

\begin{abstract}
We report the photometry of a peculiar SU UMa-type dwarf nova --- ER UMa for ten nights during 1998 December and 1999 March covering a complete rise to the supermaximum and a normal outburst cycle. Superhumps have been found during the rise to the superoutburst. A negative superhump appeared in Dec.22 light curve, while the superhump on the next night became positive and had large amplitude and distinct waveform from that of the previous night. In the normal outburst we captured, superhumps with larger or smaller amplitudes seem to always exist, although it is not necessarily true for every normal outburst. These results show great resemblance with V1159 Ori (Patterson et al. 1995). It is more likely that superhumps occasionally exist at essentially all phases of the eruption cycles of ER UMa stars, which should be considered in modeling.
\end{abstract}


\section{Introduction}
As a member of the ER UMa stars or RZ LMi stars, a small subgroup of SU UMa type dwarf novae, ER Ursae Majoris (= PG 0913+521) has received intensive attention. The history of studies for this star was described by Kato et al. (1998).

Recent photometry (Kato and Kunjaya 1995, Robertson et al. 1995) revealed that this star shows the supercycle with a period of 43 d in which the superoutburst lasts about 20 d and the normal outburst with a period of four days. These authors detected the superhumps with periods of 0.06549-0.06573 days during superoutburst. But no clear evidence of periodic hump with an amplitude larger than 0.05 mag was found during normal outburst. Subsequent observation revealed the existence of large-amplitude superhump during the earliest stage of superoutbursts (Kato et al. 1996). 

Based on the thermal-tidal instability model for SU UMa-type dwarf novae, Osaki (1995) reproduced the light curves of ER UMa by increasing mass transfer rate up to 4$\times$10$^{16}$ g s$^{-1}$.

It is rather late to establish the orbital period of ER UMa. Thorstensen et al. (1997) obtained a precise orbital period of 0.06366 d (91.67 min) based on emission line radial velocities, which provides a good condition to study in detail the variations of superhump in both super- and normal-cycles.

In this letter, we report the superhump behavior during the rise to a supermaximum and during a normal outburst.

\section{Observation}

We observed ER UMa for 44 hr over 10 nights in 1998 December and 1999 March, when the star was in a rise to its superoutburst maximum and a normal outburst respectively, using a TEK1024 CCD camera attached to the Cassegrain focus of the 1.0 m reflector at Yunnan Observatory. A total of 629 useful object frames were obtained through V filter. The exposure times are rather long, in order to assure enough signal to noise ratio even if brightness of the star drops to its minimum. The journal of the observations is summarized in Table 1.

\placetable{tbl-1}

After bias subtraction and flat-fielding, we removed the sky background and measured magnitudes of ER UMa and four secondary photometric standards (number 2, 3, 4, and 10 in the finding chart of Henden and Honeycutt 1995) so that could find the best comparison star. No variability has been detected in the differential magnitudes between number 4 and 10 (0.004 mag standard deviation) in the several-hour-observation runs, while number 2 faded by 0.1 mag in a four-hour run. We, therefore, selected number 4 as the comparison star in this study, which is 2$\arcmin$15.5$\arcsec$ southeast of ER UMa. In our differential light curves, the zero point is V = 14.2 mag and error bars for each point are, in general, less than $\pm0.017$ mag.

The AAVSO light curve (Mattei, private communication) for the system around the times of our observations is shown in Fig.1, which indicates that a rise to the superoutburst maximum and a normal outburst have been caught.

\placefigure{fig1}

\subsection{Superhumps during the rise to a supermaximum} 

Our observation in 1998 December began shortly before the minimum followed by a superoutburst and ended near the supermaximum (Fig.2). At some time on December 21, the star fell into its minimum brightness. Therefore this outburst has full amplitude of nearly 3 mag. The average rising rate is about 2.5 mag/d during Dec.21-22.

\placefigure{fig2}

The daily light curves given in Fig.3 show periodic modulations and are very different from each other in period, amplitude and waveform, which prevents us from combining them to do periodic analysis. We had to do separately for daily time series at some sacrifice of accuracy. 

ER UMa was in the decline stage of a short outburst before the superoutburst on Dec.20 and showed in its light curve an unknown origin, periodic modulation with an amplitude of about 0.1 mag. Next short time series represents the early rising stage of the superoutburst. Although we have not enough data to detect any period, periodic double-humps with unequal amplitudes of $\sim$0.3 and $\sim$0.2 mag can be seen clearly. 

\placefigure{fig3}

Evident superhump appeared in December 22 light curve with a period of 0.0589 d$\pm0.0007$ d and gradually enhance its amplitude from 0.04 mag to 0.13 mag (Table 1). The period is 7.5\% less than the orbit period, and thus the superhump is negative. On December 23, ER UMa was nearing its supermaximum. The superhump had changed to positive one with a period of 0.0654 d$\pm0.0005$ d, 2.8\% larger than the orbit period. The light curve reveals the superhump in the maximum stage to have larger amplitude from 0.21 to 0.25 mag (Table 1).

\subsection{Superhumps in the normal outburst}

Fig.4 shows the light curves obtained during 1999 March 16-21. Obviously, the period of the short outburst is about 4 days. Full amplitude of this outburst is about 2 mag. Average rising rate is 2 mag/d and average decline rate is about 0.8 mag/d.

\placefigure{fig4}

Light curves of the six days are shown in Fig.5 respectively. Marked periodic modulations can be found in March 16, 18 and 20 light curves. The periodic modulation disappeared on March 17. On the contrary, a periodic modulation was taking shape on March 21. A close inspection can marginally reveals a variation of $\sim$0.04 mag in March 19 light curve.

\placefigure{fig5}

The light curve on March 16 shows the hump with an amplitude of 0.12 mag and a period of 0.0653 d$\pm0.001$ d, 2.6\% longer than the orbital period. A periodic modulation also exists on March 18 with an amplitude larger than 0.22 mag. Period of the superhump on March 20 is only 0.0642 d$\pm 0.0004$ d, 0.88\% larger than the orbit period, and the amplitude is about 0.12 mag with gradually enhancing.
We identify the modulations (at least those occurred on March 16, 18 and 20) with superhumps for the reasons: 1) The amplitude of the modulations are large (0.12 - 0.22 mag). And 2) The period is 0.88\% - 2.6\% longer than the orbital period. To our knowledge, nothing other than superhumps can satisfy these features.

\section{Discussion}

Although we observed ER UMa for only ten nights, a complete rise to the supermaximum was caught and a normal outburst was covered. The light curves show the following features: 
1) The superhump occurred during the rise to the superoutburst. 
2) A negative superhump with only 0.07 mag amplitude appeared in December 22 light curve, while the superhump on the next night was positive and had the larger amplitude of 0.24 mag and a different waveform from that of the preceding night. and 
3) In the normal outburst we captured, the superhumps with larger or smaller amplitudes seem to always exist, although it is not necessarily true for every normal outburst.

These results show great resemblance with V1159 Ori (Patterson et al. 1995) whose light curve shows the superhump persisting far beyond the end of the superoutburst and the negative superhump appearing on two occasions. It is more likely that superhumps occasionally exist at essentially all phases of the eruption cycles of ER UMa stars.

The superhump phenomenon is now well understood to be the signature of an eccentric, precession disk (see Whitehurst 1988, Hirose, Osaki 1990). The thermal-tidal instability model (Osaki 1989) for SU UMa stars has successfully interpreted not only both normal and superoutburst behaviors but also superhump phenomenon occurring during superoutburst. In this model, the supercycle begins with a compact disk. The thermal instability produces the quasi-periodic, normal outburst, but the accretion mass in each normal outburst is less than the mass transferred from the secondary. With gradually building up of both mass and angular momentum in the disk, its radius expands with successive outburst until it eventually exceeds the critical radius for 3:1 resonance. The tidal instability leads to produce an eccentric disk. Enhanced tidal torque of the eccentric disk efficiently removes angular momentum from the disk. The last normal outburst is carried into a superoutburst. The eccentric disk exhibits a slow prograde precession, and a beat between the precession of the disk and the orbital motion of the binary is observed as a superhump. After the end of the superoutburst, the disk returns to the starting compact state. Based on this model and adopted a mass-transfer rate a factor of ten higher than that expected in the standard CV evolution theory, Osaki (1995) reproduced the light curve of ER UMa.

If we accept the fact that ER UMa (and V1159 Ori) shows superhumps at all the phases in its eruption cycle, the disk in this star must be always eccentric and precessing. In other words, the tidal torque of the eccentric disk is not strong enough to return the disk to the initial compact state. It will arouse two questions: First, what triggers the superoutburst, i.e. what mechanism causes the fast growth of tidal instability or sudden increase in the viscosity (as assumed in Murray (1998)'s simulation)? Second, if we take R$_0$ $\sim$ R$_{crit}$ in a simulation similar to that in Osaki (1995), where R$_0$ and R$_{crit}$ stand for the disk radius at the end of superoutburst and the critical disk radius for 3:1 resonance respectively, it can be inferred that the light curve of ER UMa might be reproduced with a lower mass-transfer rate. Even if so, we can not determine without concrete computation whether the value of mass-transfer rate is consistent with the well-known suggestion in which ER UMa and the other very short recurrence time systems have \.{M} in the vicinity of the critical mass-transfer rate given in the disk instability model and separating the nova-like systems from dwarf nova systems.

Another interesting problem is the explanation of the negative superhump. Patterson et al. (1993, 1997) proposed that this was the signature of the precessional motion of a tilted disk. They hypothesized that the accretion disk was simultaneously eccentric and tilted. The prograde precession of the disk's major axis gives rise to the positive superhump signal, while the retrograde precession of the disk's line of modes is responsible for the negative one. A fluid disk in a binary potential is subject to both eccentric and tilt instabilities at the 3:1 resonance (Lubow 1992). Although the three-dimensional numerical simulation by Murray and Armitage (1998) shows that the tidal inclination instability in an accretion disk is too weak to produce a significant tilt in the high state, there seems to remain room for investigating this mechanism. 

\acknowledgments

We would like to thank our referee for his inspiring comments and suggestion, Dr. Mattei for providing the unpublished AASVO data of ER UMa and Optical Astronomy Lab., Chinese Academy of Science for scheduling the observations. This work is supported by the grants 19873006 and 19733001 from the National Science Foundation of P. R. China.

\clearpage
\begin{deluxetable}{crrrrrr}
\footnotesize
\tablecaption{Summary of the observations. \label{tbl-1}}
\tablewidth{0pt}
\tablehead{
\colhead{Date} & \colhead{Start} & \colhead{Duration} & \colhead{Exposure} &\colhead{N} & \colhead{Period} &\colhead{Amplitude} \nl
\colhead{} & \colhead{(HJD 2451000+)} &\colhead{(hr)} &\colhead{(s)} &
\colhead{} & \colhead{(d)} & \colhead{(mag)}
}
\startdata

1998 Dec. 20 & 168.3187 & 3.40 & 240 & 30 &        & $\sim$ 0.1      \nl
1998 Dec. 21 & 169.3598 & 2.34 & 240 & 19 &        & $\sim$ 0.3      \nl
1998 Dec. 22 & 170.2912 & 3.94 & 240 & 54 & 0.0589 & 0.04-0.13 \nl
1998 Dec. 23 & 171.2743 & 4.39 & 240 & 54 & 0.0654 & 0.21-0.25 \nl
1999 Mar. 16 & 254.0089 & 4.69 & 240 & 49 & 0.0653 & 0.12      \nl
1999 Mar. 17 & 255.0029 & 5.15 & 240 & 57 &        & 0.15      \nl
1999 Mar. 18 & 256.0057 & 4.90 & 240 & 54 &        & 0.22      \nl
1999 Mar. 19 & 256.9979 & 5.49 & 120 & 92 &        & 0.04      \nl
1999 Mar. 20 & 258.0090 & 4.84 & 60  & 176& 0.0642 & 0.09-0.13 \nl
1999 Mar. 21 & 259.0245 & 4.33 & 240 & 44 &        & 0.38      \nl

\enddata
\end{deluxetable}

\clearpage

\figcaption[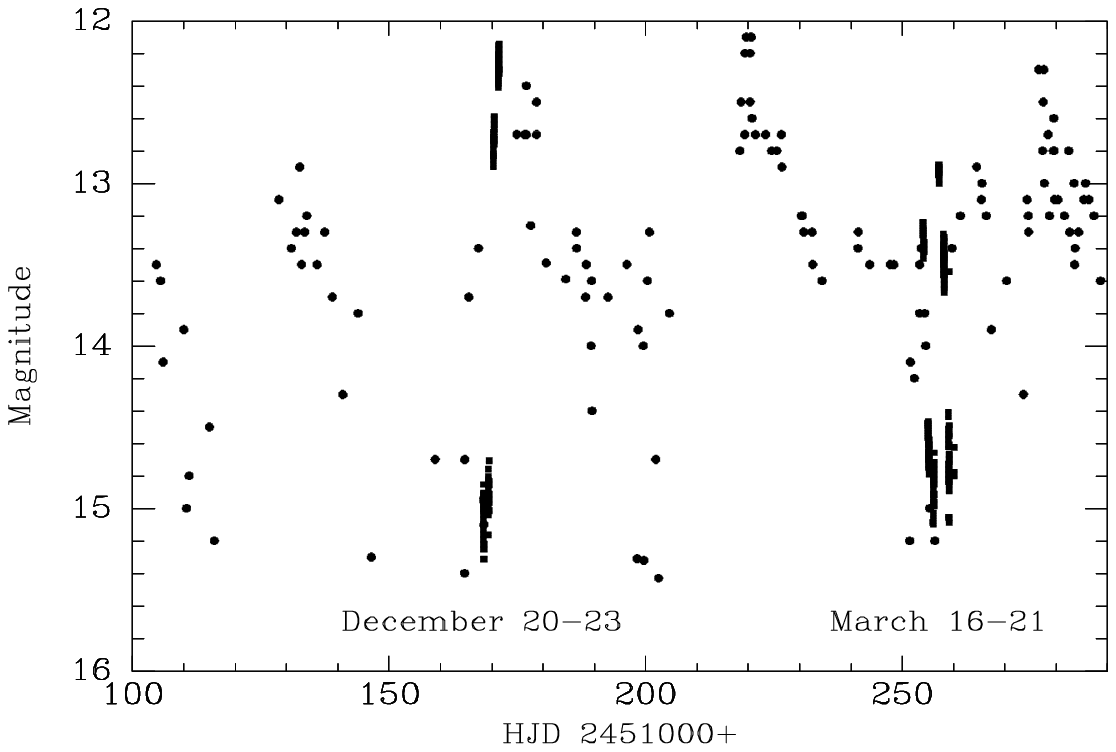]{The AASVO light curve added present observations. \label{fig1}}

\figcaption[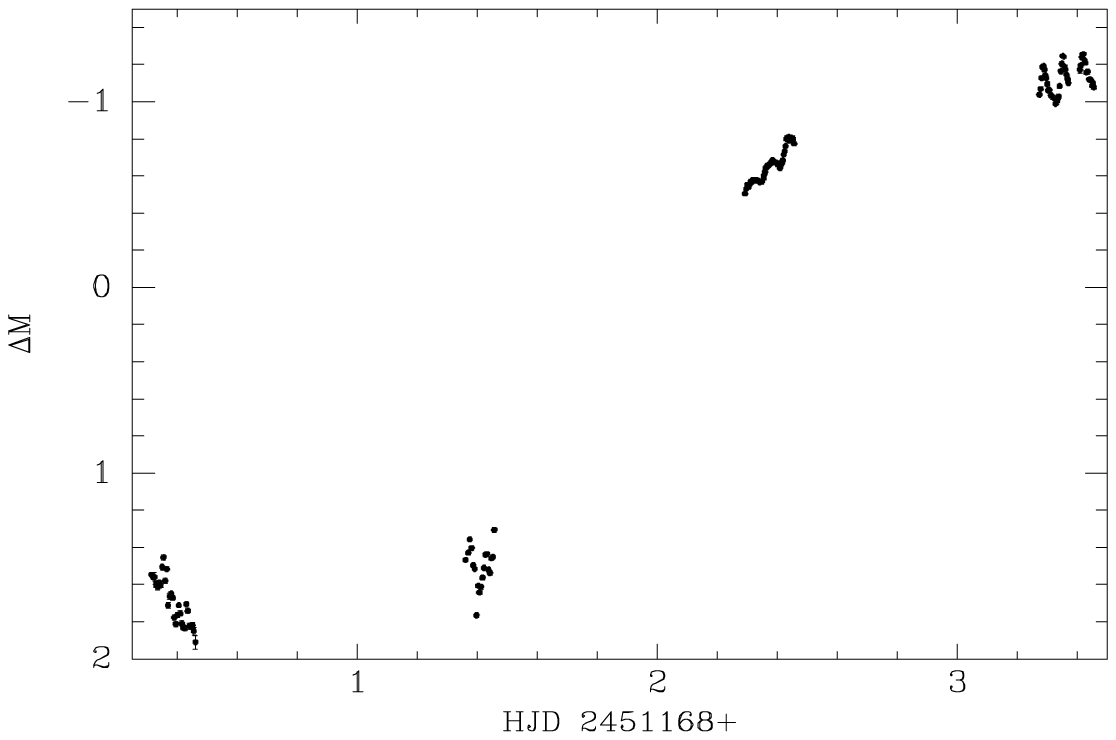]{The Observations during 1998 December 20-23, showing a complete rise to the supermaximum. Full amplitude of the superoutburst approximates 3 mag. The average rising rate is about 2.5 mag/d during Dec.21-22.\label{fig2}}

\figcaption[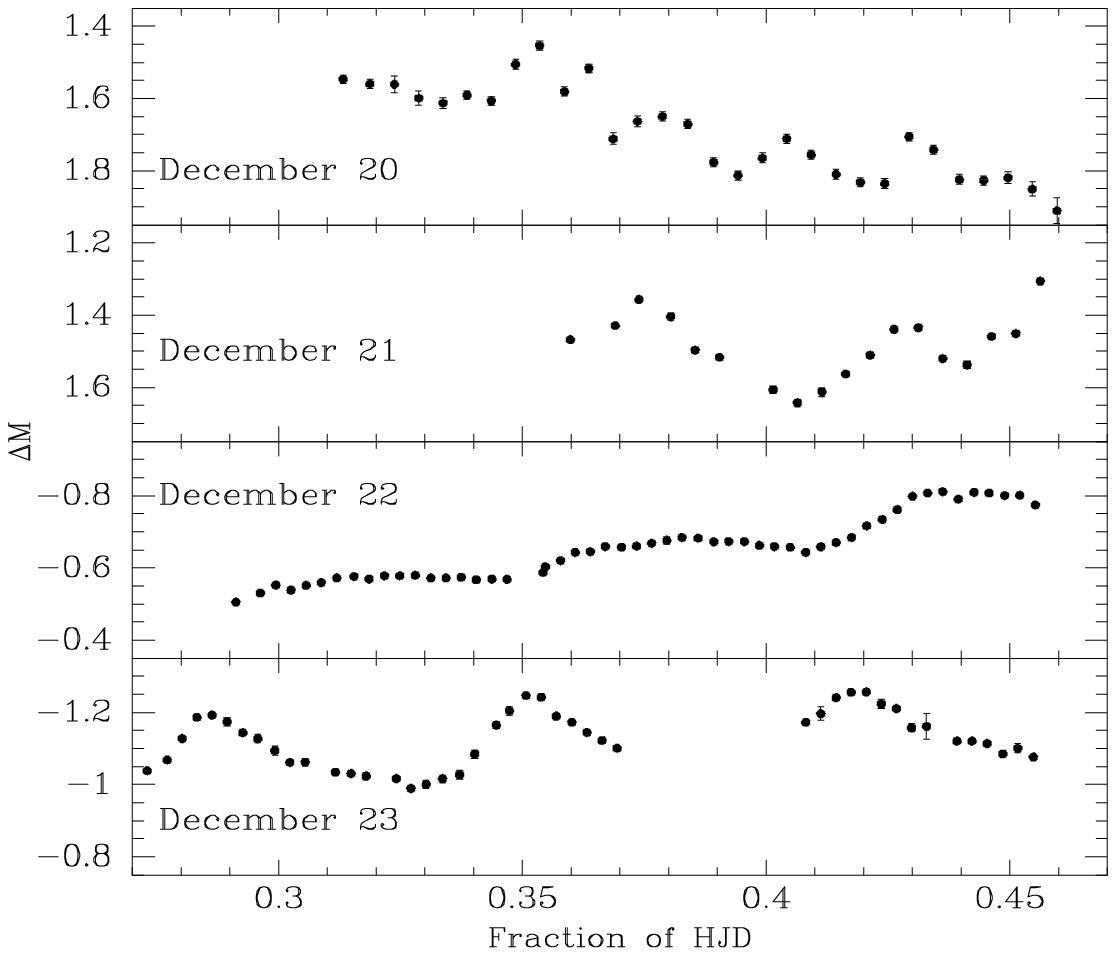]{The daily light curves with superhumps obtained in 1998 December. The superhump was negative on December 22 and changed itself to positive next night.\label{fig3}}

\figcaption[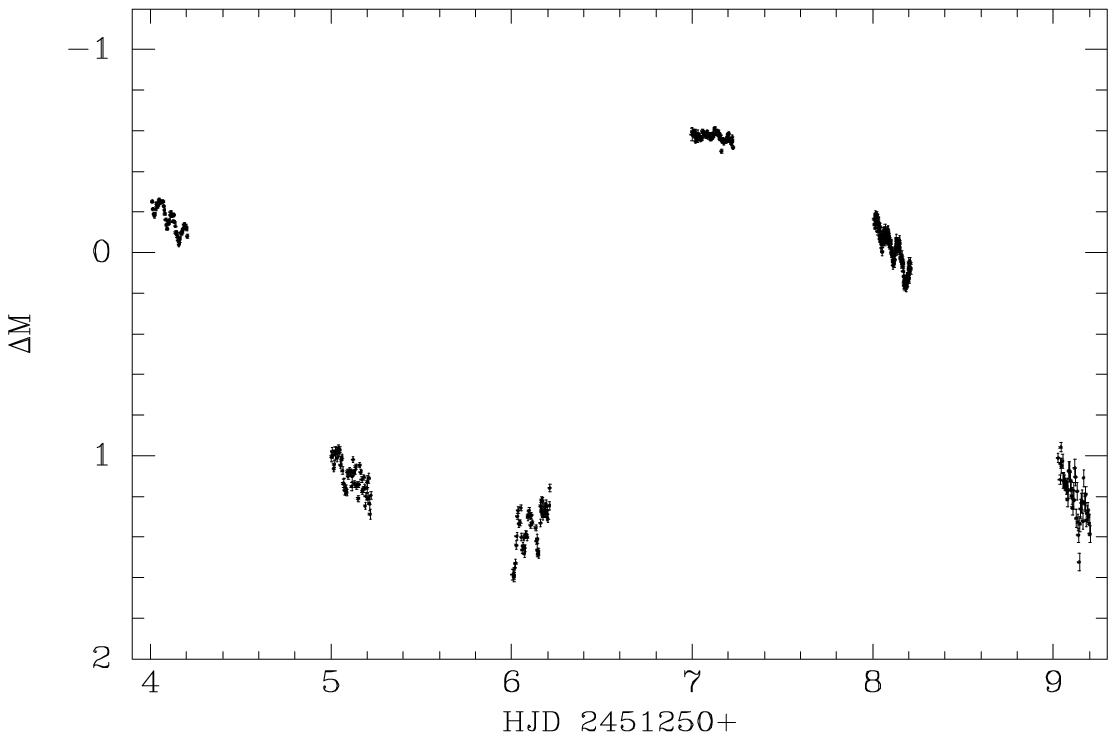]{The observations during 1999 March 16-21, showing a normal outburst with period of about 4 days. Full amplitude of this outburst is 2 mag. The average rising rate is 2 mag/d and the average decline rate is about 0.8 mag/d.\label{fig4}}

\figcaption[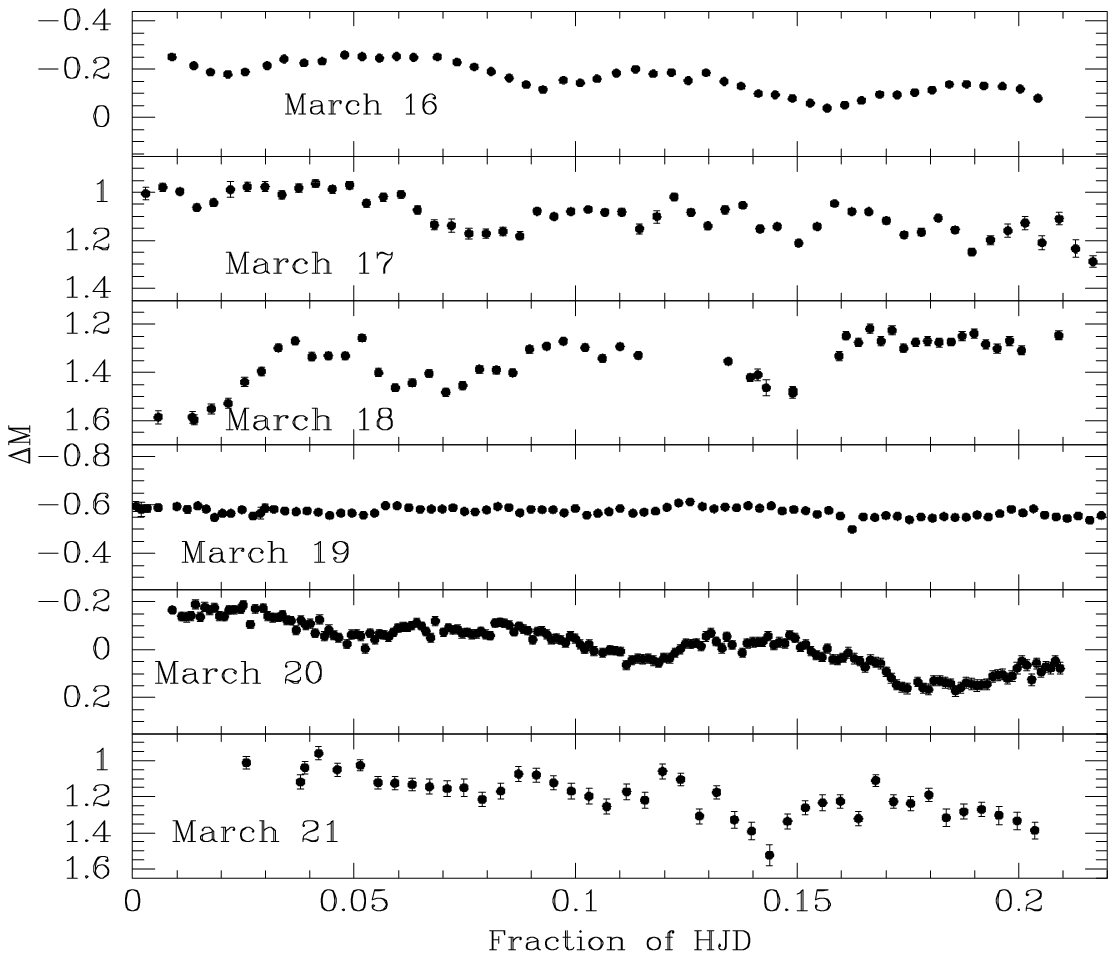]{The daily light curves obtained in 1999 March. The superhumps are visible at essentially all phases of this outburst.\label{fig5}}

\end{document}